\documentclass[aps,prb,
twocolumn,
showpacs,
superscriptaddress,
amsfonts,amssymb,amsmath,
twoside,
letterpaper,
bibnotes,
]{revtex4}

\pdfoutput=1

\usepackage{bm,color,mathrsfs,graphicx,hyperref}

\hypersetup{
    colorlinks=true, 
    linkcolor=blue,  
    citecolor=blue,  
}

\begin{document}

\title{
Interplay of magnetic and structural transitions
in Fe-based pnictide superconductors
}

\author{A. Cano}
\affiliation{
European Synchrotron Radiation Facility, 6 rue Jules Horowitz, BP 220, 38043 Grenoble, France}
\author{M. Civelli}
\affiliation{
Institut Laue-Langevin, 6 rue Jules Horowitz, BP 156, 38042 Grenoble, France}
\author{I. Eremin}
\affiliation{Institut f\"ur Theoretische Physik III, Ruhr-Universit\"at Bochum, 44801 Bochum, Germany}%
\author{I. Paul}
\affiliation{
Institut N\'{e}el, CNRS/UJF, 25 avenue des Martyrs, BP 166, 38042 Grenoble, France}

\date{\today}

\begin{abstract}
The interplay between the structural and magnetic phase transitions occurring
in the Fe-based pnictide superconductors is studied within a Ginzburg-Landau approach.
We show that the magnetoelastic coupling between the corresponding order parameters
is behind the salient features observed in the phase diagram of these systems.
This naturally explains the coincidence of transition temperatures observed in some cases
as well as the character (first or second-order) of the transitions. We also show that
magnetoelastic coupling is the key ingredient determining the collinearity of the
magnetic ordering, and we propose an experimental criterion to distinguish between a pure
elastic from a spin-nematic-driven structural transition.
\end{abstract}


\maketitle

\emph{Introduction.}--
The discovery of an unconventional high-temperature superconductivity 
in the F-doped arsenic-oxide LaFeAsO\cite{kamihara} has given rise to a great interest on iron pnictide compounds.
To date several families of Fe-based superconductors have been discovered,
and the superconducting transition temperature has been raised to above $50\,$K.\cite{Lumsden10}
All of these systems display an intriguing competition between structural, magnetic and superconducting transitions.\cite{Lumsden10,norman}
The parent compounds of the so-called 1111 and 111 families
undergo a structural transition (ST) followed by a magnetic transition (MT)
at a lower temperature,\cite{mcguire} whereas in the 122 and 11 cases
these two transitions take place simultaneously.\cite{yamazaki,wilson,FeTe}
In any case these orderings are quickly suppressed by doping or by applying pressure, which eventually gives rise to superconductivity.
The role, if any, of magnetic and elastic degrees of freedom in inducing this
superconductivity is currently an open question. There is already a growing body of
theoretical works advocating for spin fluctuation mediated
superconductivity,\cite{Mazin09} but the isotope effect observed in both magnetism
and superconductivity suggests that the elastic medium also plays a role.\cite{liu}
It is therefore compelling to understand the connection between the ST and the MT
in these systems.

The ST, for example, has been studied in Refs. \onlinecite{xu,qi}
where it has been identified with a spin-nematic ordering, but
ignoring the possible softening of the lattice itself. This idea
has been further elaborated in Ref. \onlinecite{Fernandes09},
where the observed softening of the lattice is interpreted as due
to the fluctuations of the emerging nematic degrees of freedom.
Recent studies\cite{P2-turner-china} have also associated the ST with a possible
ordering of the $d_{xy}$ and $d_{yz}$ Fe-orbitals in the Fe-As planes,
with the consequent distortions having an impact on both the magnetic order\cite{P2-turner-china,order-magnetic} and the electronic structure.\cite{P2b-superW} On the other
hand, the study of the MT in Ref. \onlinecite{Barzykin09} focuses
on the role of the magnetoelastic (ME) couplings in producing a
weak first-order MT by means of the Larkin-Pikin
mechanism.\cite{Larkin69} This latter is relevant only in the
fluctuation dominated Ginzburg regime and is subjected to the
condition of a MT with diverging specific heat in the absence of
ME coupling. Furthermore, the collinearity of the magnetic
moments is examined in Ref. \onlinecite{Eremin10} by using a
purely electronic model.

In this paper we study the ST and the MT using a Ginzburg-Landau
approach, in which the interplay between elastic and magnetic
degrees of freedom is considered explicitly. This provides a
general unified framework that goes beyond previous phenomenological
approaches and rationalizes different experimental findings. We address, in particular,
the simultaneity of the ST and MT, their character (first versus second order), the collinearity of the magnetic structure, and the spin-nematic scenario for the ST.
Our study identifies a particular ME coupling [see
Eq. \eqref{eq:ME} below] as the common key factor behind the
salient features of the ST and MT in the Fe-based superconductors.

\begin{figure}[t]
\includegraphics[width=.45\textwidth]{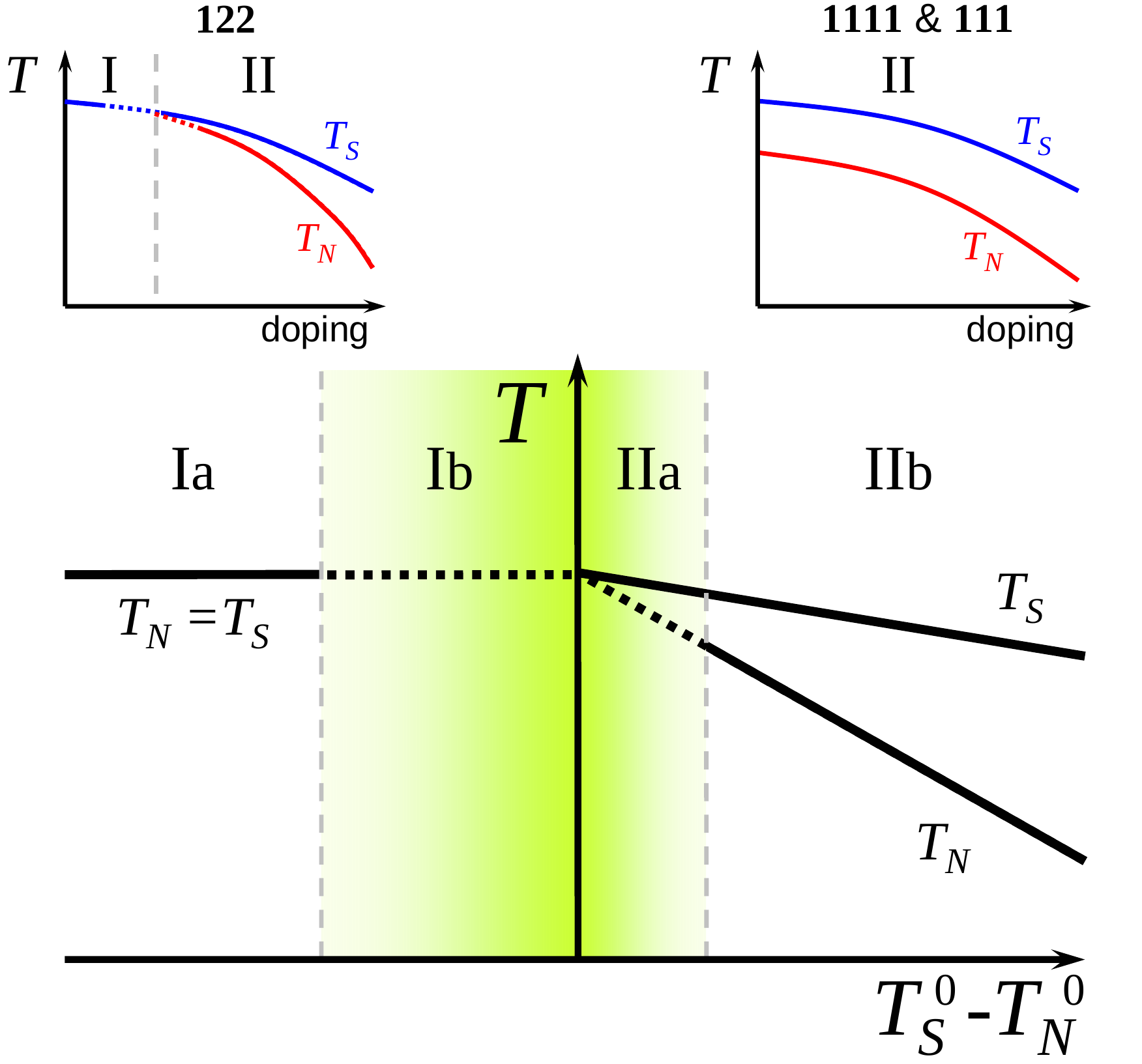}
\caption{Schematic $T$ vs. doping diagrams of the possible
structural and magnetic phase transitions in Fe-based
superconductors (top panels) and general phase diagram $T$ vs.
$T_{S}^{0}- T_{N}^{0}$ determined by our Ginzburg-Landau theory
Eq. \eqref{eq:free-energy-1} (bottom panel). 
Empirically $T_{S}^{0}- T_{N}^{0}$ increases with doping. 
In the regions I the structural and magnetic are simultaneous,
while in regions II they take place at different transition temperatures, $T_S$
and $T_N$ respectively.
Continuous (dotted) lines indicate second (first) order phase
transitions. The strength of the first order discontinuity
is indicated by the intensity of the background color.}
\label{phasediagram}
\end{figure}

Our main results are the following. (i) We derive the general
phase diagram of the ST and MT, which exhibits four qualitatively
different regimes, as shown in Fig. \ref{phasediagram}. This
explains why the simultaneous ST and MT in the 122 compounds are
sometimes observed as second-order transitions\cite{wilson}
(Ia in Fig. \ref{phasediagram}) and sometimes as first-order\cite{yamazaki}
(Ib in Fig. \ref{phasediagram}). We moreover
predict a first-order MT when this occurs separate but in the
immediate vicinity of the ST (IIa in Fig. \ref{phasediagram}).
This richness of the phase diagram is due to the ME coupling.
(ii) The collinearity of the magnetic moments is
linked to the existence of a particular ME term allowed by
symmetry. (iii) An experimental criterion is derived to
distinguish between a pure elastic from a spin-nematic-driven ST.
(iv) The fluctuations associated with ST are shown to
become critical only along certain lines of high symmetry in the
Brillouin zone. Along these lines the sound velocity vanishes in
a pure elastic ST or the spin-nematic excitations are massless in
a spin-nematic-driven ST. In both scenarios the ST has a
mean-field behavior with no divergences.\cite{qi}

\begin{figure}[t]
\includegraphics[width=.45\textwidth]{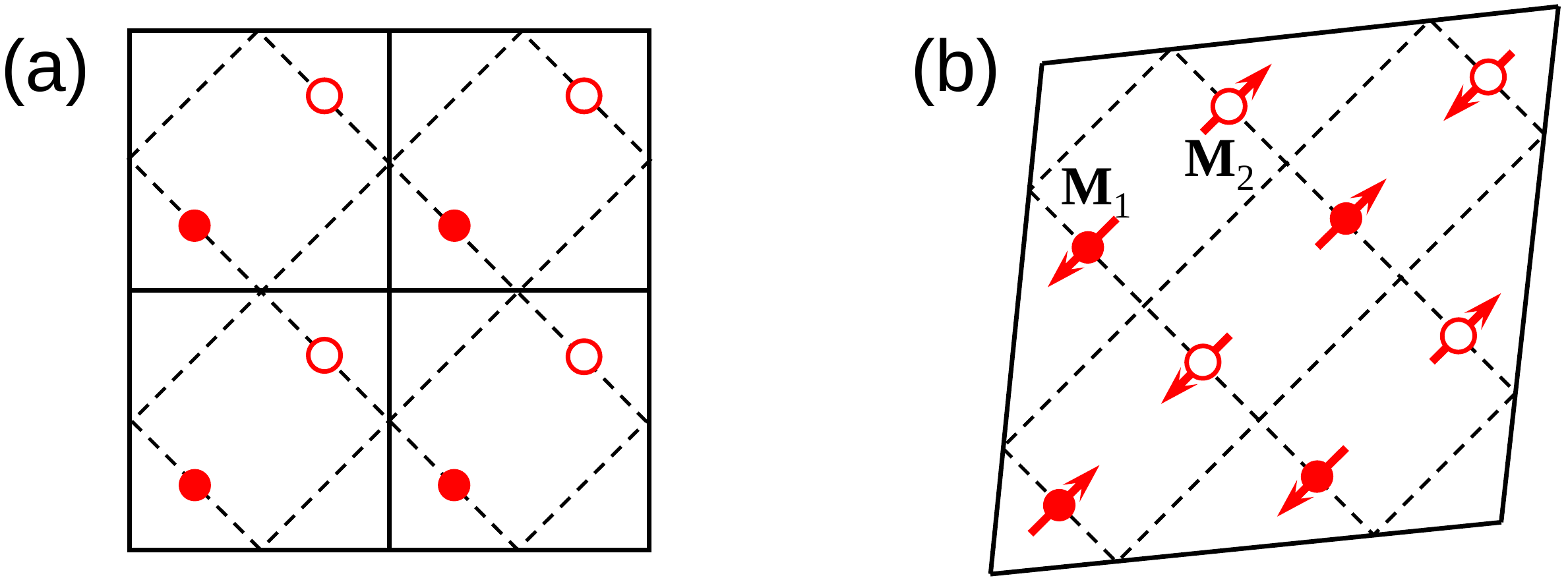}
\caption{
Tetragonal Fe-lattice (drawn as two interpenetrating sublattices) at temperatures
$T> T_S, T_N$ (a). A monoclinic distortion of the tetragonal unit cell and a
collinear N\'eel order in the two Fe sublattices take place at $T< T_S, T_N$ (b).
}
\label{distortion}
\end{figure}

\emph{Ginzburg-Landau free energy.}-- The basic ingredients to build up
a Ginzburg-Landau theory for the ST and the MT in the Fe-based
superconductors are the following.
The initial order parameter for the ST can be taken as the shear strain
$u_{xy}$,\cite{footnote0} where subscripts refer to the principal axes of
the tetragonal lattice (with two and four Fe atoms per unit cell for the 1111 and the
122 systems respectively, see Fig. \ref{distortion}a). Thus $u_{xy}\not =0 $
implies the monoclinic distortion of the tetragonal unit cell observed experimentally,
as illustrated in Fig. \ref{distortion}b. For the case of a spin-nematic driven ST
an additional order parameter $\sigma$ is required, which will be explained in more
detail below. On the other hand the description of the MT requires two N\'{e}el
vectors,\cite{footnote1} $\mathbf L_1$ and $\mathbf L_2$, which can be associated with
the magnetizations $\mathbf M_1$ and $\mathbf M_2$ of the two inter-penetrating Fe
sublattices (Fig. \ref{distortion}b). The Ginzburg-Landau free energy then can be
conveniently written as
\begin{equation}
\label{eq:free-energy-1}
F_{GL} = F_{M} + F_{E} + F_{ME}.
\end{equation}
Here the magnetic part is
$F_{M} = {1\over 2}A( L_1^2 +  L_2^2) + {1\over 4}B (L_1^4 +  L_2^4)
$.
In principle fourth order terms of the form $L_1 ^2 L_2 ^2$ and
$(\mathbf L_1 \cdot \mathbf L_2)^2$ are also allowed by symmetry, but we omit them
because the former does not affect the results qualitatively while the latter is
generated by the ME coupling (see below). For the elastic part we consider explicitly
the critical strain $u_{xy}$ only and we write
$F_E = {1\over 2}c_{66} u_{xy}^2 + {1\over 4}\beta u_{xy}^4$.
The coefficients $A$ and $c_{66}$ are
assumed to vary with the temperature as $A= A^{\prime}(T-T_N^0)$ and
$c_{66} = c_{66}^{\prime}(T-T_S^0)$. $T_N^0$ and $T_S^0$ are the nominal MT and ST
temperatures respectively (not to be confused with the actual transition temperatures
$T_N$ and $T_S$), taken as the control parameters of our theory. 
It is our empirical obsevation that the results of the
Landau-Ginzburg theory are consistent with the current experiments
if we assume that $T_S^0 - T_N^0$ increses with doping. 
A possible expalantion of this trend at very low doping is the following. It is
likely that the effect of doping is more on the magnetic sector 
(say, via the loss of nesting between electron and hole bands)
than the elastic medium which it affects indirectly. Thus, it may not 
be unreasonable to expect that at low doping $T_S^0$ stays relatively 
unchanged while $T_N^0$ decreases (since the system hardens magnetically).
The remaining coefficients $B$, $C$ and $\beta$ are taken as positive constants.
As regards the ME term, the key contribution is\cite{qi,Fernandes09,Barzykin09}
\begin{equation} \label{eq:ME}
F_{ME} = g_1 u_{xy}(\mathbf L_1 \cdot \mathbf L_2),
\end{equation}
where $g_1 >0$ in order to be consistent with the experimental observation that the
ferromagnetic Fe-Fe bonds are shorter than the antiferromagnetic ones in the
collinear-N\'{e}el state. To our purposes, the standard magneto-striction
$u_{ll}( L_1^2 +  L_2^2)$ can be neglected because it does not change
qualitatively the results. It is worth mentioning that, in the context of
the cuprate superconductors, a similar free energy $F_{GL}$ was used to study
the structural transition of doped La$_2$SrCuO$_4$.\cite{sarrao}

\emph{Phase diagram.}-- We are in a position to discuss the general phase diagram
for the ST and MT in the Fe-based superconductors (Fig. \ref{phasediagram}).
We anticipate that, experimentally, undoped and lightly doped 122 systems fall in regimes Ia and Ib with simultaneous transitions of  second and first order respectively. On the other hand, the 1111 and 111 compounds appear to fall in the regime IIb, where the transitions are separate ($T_S> T_N$) and second order. In the regime IIa the ST is second order while the separate MT is first order. This regime has not yet been reported experimentally, perhaps because it is narrow in the tuning of physical parameters (such as pressure and doping) and therefore difficult to detect.

Within our framework, the fact that these transitions sometimes are observed separate and sometimes simultaneous is explained as follows. Note that, once the magnetic order sets in and the collinear-N\'{e}el state is formed, there is an effective shear stress that produces a monoclinic distortion via the ME term
(Eq. \ref{eq:ME}). Thus, the MT in these systems implies a ST, while the converse is evidently not true.
This gives rise to the regimes I and II in the phase diagram shown in Fig. \ref{phasediagram}, which are defined by the conditions $T_S^{0} < T_N^{0} $ and $T_S^{0} > T_N^{0} $ respectively.
Thus, as the temperature is lowered in regime II there is first the ST at $T = T_S^0$, where $c_{66} = 0$, followed by a separate MT at $T_N = T_N^0 + [g_1 u_0(T_N)]/A^{\prime}$, where
$u_0(T) = \pm[(T_S^0 - T) c_{66}^{\prime}/\beta]^{1/2}$ is the value of the strain order parameter
$u_{xy}$ in the monoclinic phase at $T < T_S$.
In the regime I, however, the system encounters the MT first at $T = T_N^0$, where $A = 0$,
but simultaneously the ST occurs due to the effective stress in the ME term.
In this regime, unlike in II, there is no instability of the lattice because the elastic modulus $c_{66}$ stays finite (and jumps across the transition).

These two regimes are divided in their turn into subregimes Ia, Ib, IIa and IIb according to the character, first or second order, of the MT.
This is again due to the ME coupling, and has nothing to do with symmetry reasons (such as the existence of cubic invariants).
The point is that the effective coefficient of fourth-order term in the magnetic sector can become negative due to the ME coupling (in which case, one has to include higher order terms to bound the free energy from below).
This is readily seen in the regime I ($T_S^{0} < T_N^{0} $), where the elastic sector
is stable and one can ignore the $u_{xy}^4$ term in $F_{E}$.
Indeed minimizing $F_{GL}$ one gets $u_{xy} = - g_1 (\mathbf L_1 \cdot \mathbf L_2)/c_{66}$, which implies that the elastic sector generates an effective magnetic term of the form $- g_1^2 (\mathbf L_1 \cdot \mathbf L_2)^2/(2 c_{66})$.
The MT into the collinear-N\'{e}el state turns first order when this term is
sufficiently strong, with the tri-critical point given by $c_{66} = g_1^2/B$ and $A =0$.
The logic is quite similar
in the regime II ($T_S^{0} > T_N^{0} $), where one first has to expand
$u_{xy} = u_0 + \delta u_0$ in the monoclinic phase to then obtain
(after further minimization) $\delta u_0= - g_1
(\mathbf L_1 \cdot \mathbf L_2)/(2|c_{66}|)$. This implies an effective magnetic
term $- g_1^2 (\mathbf L_1 \cdot \mathbf L_2)^2/(4 |c_{66}|)$ which drives the MT
first order, with the tri-critical point given by $A= g_1 u_0(T_N)$ and
$c_{66} = - g_1^2/(2B)$.
These results have to be contrasted to Ref. \onlinecite{Barzykin09}, where strong fluctuations
are required and weaker first-order transitions, with narrower hysteresis, are expected.
At the mean field level, it is the vicinity of the structural instability
(where $c_{66}=0$) to determine the area (region IIa in the phase diagram of Fig.
\ref{phasediagram}) where the MT becomes first order due to the ME coupling. In fact
the discontinuity is the stronger the closer is to the point
$T_S^0 =T_N^0$ (as indicated by the color intensity in Fig. \ref{phasediagram}).

\emph{Collinearity of the magnetic order.}---
At this stage, the role of the ME coupling in favoring the collinear-N\'{e}el state
observed experimentally\cite{cruz} is quite straightforward. In regime II the collinearity is
due to the term $g_1 u_0(\mathbf L_1 \cdot \mathbf L_2)$, which lifts the degeneracy
of the transition temperatures for
$\mathbf L_\pm= {1\over \sqrt{2}}(\mathbf L_1 \pm \mathbf L_2)$. This
implies that only one between $\mathbf L_\pm$ is non-zero in the ordered state (which one
in particular depends on the sign of $u_0$, i.e., on the ferroelastic domain).
In regime I, however, the reason is different. In this case the set up of the
collinear order is associated with the negative coefficient $- g_1^2/(2 c_{66})$
in front of the term $(\mathbf L_1 \cdot \mathbf L_2)^2$, generated by the ME coupling.
Presumably, the vicinity to the structural instability makes this ME term
strong enough to overcome any other contribution (due to, e.g., magnetic interactions)
and thus forces the collinearity.

\emph{Spin nematic versus pure elastic instability.}---
In the region II of the phase diagram, where the two transitions are separate, there are
two different physical scenarios to conceptualize the ST: a pure ferroelastic transition
or a transition induced by a spin-nematic ordering. In the following we discuss some
physical implications that may allow to determine experimentally which one of the two
scenario is realized in the Fe-based superconductors.

(a)
In the first scenario we have a proper ferroelastic instability in $F_E$ caused by the
vanishing of the elastic modulus $c_{66}$ (as described above). An immediate consequence
is that the temperature range over which the monoclinic distortion
has square-root behavior $u_0(T) \propto (T_S -T)^{1/2}$ is the same as the temperature
range over which the elastic modulus is expected to have a linear dependence
$c_{66} \propto |T - T_S|$.

(b)
The second scenario is more subtle, as it arises from the possibility that,
above $T_N$, the system enters a spin nematic state in which the two N\'{e}el
vectors are zero in average but fluctuate in phase,
i.e., $\sigma \equiv \langle \mathbf L_1 \cdot \mathbf L_2 \rangle \not =0$.\cite{premi}
In this case the system is expected to undergo a pseudo-proper ferroelastic transition
driven by the nematic ordering. This scenario can be described by the free energy
\begin{equation}
\widetilde{F}_{GL} = {a \over 2} \sigma^2 + {b \over 4} \sigma^4 + {c_{66} \over 2} u_{xy}^2
+ {\beta \over 4} u_{xy}^4 + g_2 u_{xy} \sigma,
\end{equation}
where $a = a^{\prime}(T-T_n^0)$ with $T_n^0$ being the nominal nematic transition
temperature, while all other coefficients are $T$-independent (including $c_{66}$ for
this discussion). The structural-nematic transition, below which
$u_{xy} \propto \sigma \neq 0$, occurs at $T_S = T_n^0 + g_2^2/(c_{66}a^{\prime})$ where
the effective elastic modulus $\widetilde{c}_{66} = c_{66} - g_2^2/a$ vanishes. In this
case, the range of temperatures over which $\widetilde{c}_{66} \propto |T-T_S|$ is
restricted to the condition $|T-T_S| < g_2^2/(c_{66}a^{\prime})$. This range is
in principle different from the $T$-range over which the monoclinic distortion has
square-root dependence, as we have discussed in point (a) above.

\begin{figure}[t]
\includegraphics[width=.45\textwidth]{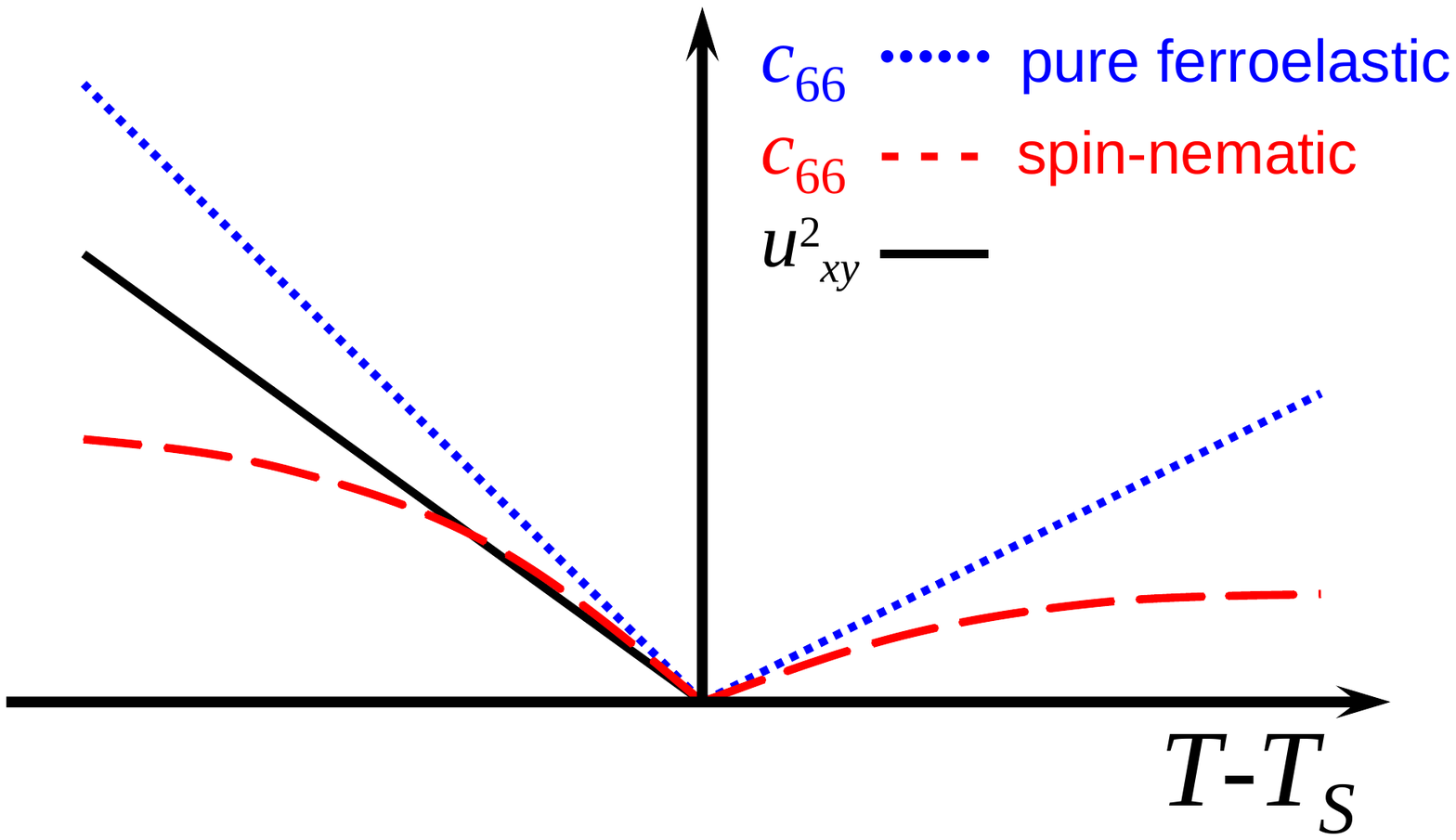}
\caption{
Expected dependence on temperature of the squared spontaneous
strain $u_{xy}^2$ and the elastic modulus $c_{66}$ in scenarios (a)
proper ferroelastic instability and (b) driven by spin-nematic ordering.
Large deviations from a linear behavior in $c_{66}$ as compared to $u_{xy}^2$
indicate spin-nematic ordering.}
\label{shearmodulus}
\end{figure}

This simple observation, which is schematically represent in Fig. \ref{shearmodulus},
can be used as a quantitative criterion to distinguish between the pure ferroelastic and
spin-nematic-induced scenarios. These considerations are based on a mean field analysis
of the ST, which we justify in the following.

\emph{Mean-field behavior for the structural transition.}--
Previous studies on ferroelastics\cite{cowley} have shown that a second-order ST,
like in the proper ferroelastic scenario (a), displays a mean-field behavior. This
follows from the fact that at the transition point, where the elastic modulus
$c_{66} = 0$, the phonon velocity remains finite everywhere in the Brillouin zone
($\mathbf q$-space), except along the two lines of high symmetry $q_x = 0$ and $q_y =0$
on the $q_z =0$ plane. Consequently, except for these ``soft" lines, the long-wavelength
critical excitations associated with the strain field $u_{xy}$ are gapped and
the transition in three dimensions is mean-field like.\cite{cowley} The contribution
of these excitations to $F_{E}$ can be written as
$\delta F_{E} = {1\over 2}\sum_{\mathbf q \not = 0 }
c_{66} (\mathbf q) |u_{xy}(\mathbf q)|^2 + \cdots$,
where the ellipsis denotes non-critical modes and interaction terms,
\begin{equation}
c_{66} (\mathbf q) \approx c_{66} + c_1 \cos^2\theta + c_2\sin^2 2\varphi \sin^4 \theta + \kappa q^2
\end{equation}
near the soft lines, and $\theta$ and $\varphi$ represent the polar and the azimuthal
angles respectively in $\mathbf q$-space.\cite{cowley} The $\kappa q^2$ contribution is
due to harmonic terms in the free energy with higher order derivatives (the usual
stiffness associated with collective excitations) and $c_1$ and $c_2$ are constants that
depend on the elastic moduli of the system.\cite{footnote2}
In the scenario (b) the situation is slightly more subtle. Indeed, if we ignore the
coupling $g_2 u_{xy} \sigma$, it appears that the nematic transition, which belongs to
the Ising class, is in three dimensions below its upper critical dimension and
therefore it is not mean-field type. However, the ME coupling gives
rise to anisotropic corrections to the mass of the $\sigma$-field.
In fact the resulting mass term can be written as
$F_{\sigma} = {1\over 2}\sum_{\mathbf q \neq 0} a (\mathbf q) | \sigma (\mathbf q) |^2$,
where $a(\mathbf q) = a - g_2^2/c_{66}(\mathbf q)$. As a consequence, in this scenario
too, the long-wavelength critical excitations are gapped except along two soft lines.
\cite{Levanyuk70} So the ST is completely mean field in any case.

We finally discuss some limitations of our study. We have
restricted ourselves to magnetic states describable in terms of
the N\'{e}el vectors $\mathbf L_1$ and $\mathbf L_2$. This includes the
collinear $(\pi,0)$ [or $(0,\pi)$] ``stripe'' order observed
experimentally. In the case of a different magnetic ground state
the corresponding Ginzburg-Landau theory may have a different
form, with structural and magnetic orderings not necessarily
coupled. This seems to be the case in BaMn$_2$As$_2$, where a
G-type antiferromagnetic order is found and no structural
distortion is observed. \cite{singh09} The same applies to 11
Fe-chalcogenides, for which, in addition, the question of whether
the magnetic order is intrinsic or arises from non-stoichiometry,
i.e., due to interstitial Fe-ions inducing weak charge
localization, is still open.\cite{liu10} Critical fluctuations
are also not considered in our theory. While we expect that they
will not change the nature of the ST, they certainly will enhance
the effects of the Larkin-Pikin mechanism\cite{Larkin69} on the
MT. We do expect therefore an enlargement of the first order MT
region (Ib and IIa in Fig. \ref{phasediagram}). All these
considerations deserve deeper investigation in future studies.

In summary, we have shown that the magneto-elastic coupling Eq. \eqref{eq:ME} 
plays a key role in producing the specific features of the structural and magnetic
transitions in the Fe-based superconductors. This coupling is responsible
for the simultaneity of the transitions observed in the 122 systems and explains
the different characters (first-order and second-order) of the transitions reported
so far. It also naturally explains the collinearity of the magnetic order.
We have also addressed the question of the possibility of a spin-nematic driven structural
transition, indicating an experimental way to discriminate this type of scenario from
a proper ferroelastic transition.

We acknowledge G. Garbarino, E. Kats,
M. N\'{u}\~{n}ez-Regueiro, J. Schmalian and T. Ziman for stimulating discussions.


\begin{references}
\bibitem{footnote0} The strain tensor is defined as $u_{ij} = {1\over 2}(\partial_i u_j + \partial_j u_i)$, where $\mathbf u$ is the displacement vector in a continuum medium description.
\bibitem{footnote1}
The two sublattice magnetizations associated with the $(\pi,0)$ [or $(0,\pi)$] orderings
observed experimentally can be defined as
$\mathbf M_1= \mathbf L_1\cos (\mathbf Q \cdot \mathbf R_1) \cos (\pi z)$ and
$\mathbf M_2= \mathbf L_2\cos (\mathbf Q \cdot \mathbf R_2)\cos (\pi z)$,
where $\mathbf Q= (\pi, \pi)$ and $\mathbf R_{1(2)}$ are points on the Fe sublattice 1(2). The observed moments are $\sim 0.3-0.7\mu_B$ in the 1111 and 111 and
$\sim 1.5\mu_B$ in the 122 compounds.
\bibitem{footnote2}
If, for example, $F_{E} = {\lambda\over 2} u_{ll}^2 + \mu u_{ij}^2
+ {1\over 2}(c_{66} - 4 \mu) u_{xy}^2$ where $\lambda$
and $\mu$ are the Lame coefficients, one has $c_1 = 4 \mu$
and $c_2 = 16 \mu (\lambda + \mu)/(\lambda + 2 \mu)$.
The most general form of $F_{E}$ with tetragonal symmetry involves six elastic constants,
but, for convenience's sake, we use a simplified model with only three elastic constants.
\end{references}
\end{document}